\documentclass[amsmath,amssymb,preprint]{revtex4-1}

\usepackage{graphicx}
\usepackage{dcolumn}
\usepackage{bm}

\begin{document}

\title{The deconfined phase of ${\cal N}=1$ SUSY Yang-Mills: bound states and the equation of state}

\author{Gwendolyn \surname{Lacroix}} 
\email[E-mail: ]{gwendolyn.lacroix@umons.ac.be}
\author{Claude \surname{Semay}}
\email[E-mail: ]{claude.semay@umons.ac.be}
\affiliation{Service de Physique Nucl\'{e}aire et Subnucl\'eaire,
Universit\'{e} de Mons -- UMONS, Place du Parc 20, 7000 Mons, Belgium}

\author{Fabien \surname{Buisseret}}
\email[E-mail: ]{fabien.buisseret@umons.ac.be}
\affiliation{Service de Physique Nucl\'{e}aire et Subnucl\'{e}aire,
Universit\'{e} de Mons -- UMONS,
Place du Parc 20, 7000 Mons, Belgium;\\ 
Haute \' Ecole Louvain en Hainaut (HELHa), Chauss\'ee de Binche 159, 7000 Mons, Belgium}

\date{\today}

\begin{abstract}
The properties of the deconfined phase of ${\cal N}=1$ supersymmetric Yang-Mills theory in $(3+1)$-dimensions are studied within a $\cal T$-matrix formulation of statistical mechanics in which the medium under study is seen as a gas of quasigluons and quasigluinos interacting nonperturbatively. Emphasis is put on the temperature range (1-5)~$T_c$, where the interaction are expected to be strong enough to generate bound states. Binary bound states of gluons and gluinos are indeed found to be bound up to 1.4 $T_c$ for any gauge group. The equation of state is given for SU($N$) and $G_2$; it is found to be nearly independent of the gauge group and very close to that of non-supersymmetric Yang-Mills when normalized to the Stefan-Boltzmann pressure and expressed as a function of $T/T_c$. Finally the orientifold equivalence is shown to hold at the level of the equation of state and its accuracy at $N=3$ is shown to be very good. 
\end{abstract}


\maketitle

\section{Introduction}

A full understanding of the phase diagram of quantum chromodynamics (QCD) is a major goal in the field. As such, it has been the subject of intense investigation, see \textit{e.g.} \cite{yagi}. Apart from QCD, the finite-temperature behavior of generic Yang-Mills (YM) theories -- with arbitrary gauge groups and/or matter in higher representations -- is a topic that appears no less challenging, but about which less information is available. A key result is that a phase transition from confinement to deconfinement seems to be a generic feature of ordinary YM theories; at least it has been observed with gauge groups G$_2$, SU($N>3$), Sp(2) and E$_7$  \cite{G2,braun,Dumitru:2012fw}. Moreover the equation of state (EoS) above the deconfining temperature ($T_c$) is nearly independent of the gauge group (at least for SU($N$) and G$_2$) once normalized to the Stefan-Boltzmann pressure and expressed as a function of $T/T_c$ \cite{Dumitru:2012fw,LSCB,panero}. 

Among the possible couplings of pure YM theory to matter, an appealing one is the inclusion of a Majorana fermion in the adjoint representation of the gauge group, leading to the $\mathcal{N} = 1$ supersymmetric (SUSY) YM theory \cite{salam}, the adjoint quarks being called the gluinos. The $\beta$-function of this theory has been exactly computed from instanton calculus \cite{Novi83}, and reads $\beta(g)=-\frac{g^3}{16\pi^2}\frac{3N}{1-\frac{g^2 N}{8\pi^2}}$ with the gauge group SU($N$). As in the pure YM case, it suggests both asymptotic freedom and confinement. The $\mathcal{N} = 1$ SUSY YM bound state spectrum at zero temperature has been investigated by resorting to effective actions \cite{Farrar:1997fn} and to lattice computations \cite{Bergner:2013nwa}. At finite $T$, this theory is expected to exhibit a deconfining phase transition: Recent results indicate that it might be the case for any gauge group \cite{Bergner:2014saa,Anber:2014lba}. At very high temperatures, the deconfined phase should behave as a conformal gas of gluons and gluinos \cite{Amat88}. A peculiar feature of the SU($N$) $\mathcal{N} = 1$ SUSY YM is that it is equivalent to one-flavor QCD at large $N$ provided that quarks are in the two-indices antisymmetric representation of SU($N$), which is isomorphic to the fundamental one at $N=3$. This equivalence is called orientifold equivalence and has attracted a lot of attention since the work \cite{Armo03}. 

We present here a first study of the thermodynamic features of the deconfined phase of $\mathcal{N} = 1$ SUSY YM, including the existence (or not) of bound states and the EoS. Such results have, to our knowledge, never been obtained.  For the sake of clarity, we will discuss our main results and summarize our formalism, while we refer the interested reader to \cite{LSCB,SUSYlong} for technical details concerning the computations performed. 


\section{The model}\label{secmod}

Our main assumption is that the deconfined phase of the $\mathcal{N} = 1$ SUSY YM theory can be described as a relativistic non ideal gas of transverse quasigluons and quasigluinos (the effective degrees of freedom propagating in the medium) in which two-body interactions are dominant. It is actually an extension to a supersymmetric case and an arbitrary gauge group of the picture developed in the celebrated paper \cite{Shuryak:2004tx}. Moreover, the quasiparticle properties and their two-body interactions will be investigated by resorting to a $\cal T$-matrix formulation similar to that of \cite{tmat}, in which valuable results about heavy quark flavors in the quark-gluon plasma have been obtained. We define the mass of a quasiparticle ($m$) as the sum of a bare mass ($m_0$) and a thermal mass ($\propto \kappa(T)$). The gauge-group dependence of this latter is chosen to be the one obtained in Hard-Thermal-Loop computations \cite{HTL}. So, for a quasiparticle in the representation $r$ of the gauge group,
\begin{equation}\label{m2}
m^2(T)=m_0^2+\frac{C^{r}_2}{C^{adj}_2}\kappa^2(T),
\end{equation}
where $C^r_2$ is the quadratic Casimir of the gauge group in the representation $r$ ($adj$ is the adjoint representation). The interaction potential between quasiparticles $i$ and $j$ is assumed to have the one-gluon-exchange form 
\begin{equation}
V(r,T)=\frac{\vec M_i\cdot \vec M_j}{C^{adj}_2} v(r,T),
\end{equation}
where $\vec M_i$ is the generator of the gauge algebra in the representation $i$.  

The EoS can then be computed by resorting to $S$-matrix formulation of statistical mechanics proposed in \cite{dashen} according to which the grand potential of an interacting relativistic particle gas, $\Omega$, expressed as an energy density, is given at zero chemical potential by 
\begin{eqnarray}\label{pot0}
\Omega&=&\Omega_0+\sum_\nu\bigg[\Omega_\nu\nonumber \\
&-&\frac{1}{2\pi^2\beta^2}\int^\infty_{M_\nu} \frac{d\epsilon}{4\pi i} \epsilon^2  K_2(\beta\epsilon) \left. {\rm Tr}_\nu \left({\cal S}S^{-1}\overleftrightarrow{\partial_\epsilon}S \right)\right|_c\bigg].
\end{eqnarray}  
$\Omega_0$ is the grand potential of the free relativistic quasiparticle gas, while the second term accounts for interactions in the medium and is a sum running on all the quantum numbers $\nu$ needed to label a channel ($M_\nu$ is the sum of the particle masses in the channel $\nu$). The sum $\sum_\nu$ reads here $\sum_{J^{PC}}\sum_{{\cal C}}$, where ${\cal C}$ is the color channel, and $J^{PC}$ is the spin/parity channel (labels $C$ and $P$ must be dropped off if they are undefined). As in \cite{LSCB}, we only consider two-body channels, expected to be the dominant ones.  $\Omega_\nu$ is the contribution coming from bound states in a given channel while the last term is the scattering term above the threshold. It is a function of the $S$-matrix, of the symmetrizer ${\cal S}$ if needed and of $\beta = 1/T$. The subscript $c$ means that only the connected scattering diagrams are taken into account. Note that $A\overleftrightarrow{\partial_x} B=A(\partial_xB)-(\partial_xA)B$.

By definition, $S=1-2\pi i\, \delta(\epsilon-H_0)\, {\cal T}$, where ${\cal T}$ is the off-shell $\cal T$-matrix and where $H_0$ is the free Hamiltonian. $G_0$ being the free propagator, ${\cal T}$ is obtained by solving the Lippmann-Schwinger equation ${\cal T}=V+ V\, G_0\, {\cal T}$ thanks to the Haftel-Tabakin algorithm \cite{haftel}. Note that in-medium effects have been included at the level of the propagator according to the prescription of \cite{Prat94}. The $\cal T$-matrix and the potential $V$ are understood to be in a given two-body channel; the basis states needed for the matrix elements are computed within the helicity formalism of \cite{jaco} in order to handle transverse particles. In Eq.~(\ref{pot0}), a channel is included only if its cross section (based on a ${\cal T}$-matrix calculation) is at least 25\% of the cross section of the channel involving the same species with the lowest possible $J$ \cite{SUSYlong}. 

Once the $\cal T$-matrix is known in all the channels taken into account, the pressure and trace anomaly are simply given by
\begin{equation}\label{pta}
\frac{p}{p_{SB}}=-\Omega \quad {\rm and}\quad \frac{\Delta}{p_{SB}}=-\beta\,\partial_\beta\left(\frac{p}{p_{SB}}\right),
\end{equation}
where, for better convenience, these quantities are normalized to their corresponding Stefan-Boltzman pressure $p_{SB} = - \displaystyle\lim_{m \rightarrow 0} \Omega_0$.

\section{Parameters}\label{secparam}
As numerical input, we take the lattice data of \cite{latpot}, giving the free energy $F_1(r,T)$ of static quark-antiquark pair bound in color singlet in quenched SU(3) QCD. From those data, the internal energy $U_1(r,T)$ can be computed. According to the arguments given in \cite{tmat}, the gauge-group independent part of the interaction potential is then $v(r,T)=4(U_1(\infty,T)-U_1(r,T))/9$, while the gauge-group independent part of the thermal mass is taken to be $\kappa(T)=3 U_1(\infty,T)/4$. The asymptotic part of the internal energy can indeed be interpreted as a contribution from the noninteracting sources, hence as a self-energy term. 

The meaningful parameters are the ratios $T/T_c$, $m_0/\sqrt \sigma$ and $T_c/\sqrt \sigma$, where $\sigma$ is the fundamental string tension. For the computations, we take $\sigma=0.176$~GeV$^2$ as in \cite{LSCB,SUSYlong}. In the non-SUSY YM case, identifying the critical temperature to the Hagedorn temperature of a bosonic closed string theory in $(3+1)$-dimensions agrees well with currently known lattice data \cite{hage,hage2}: $T_c(\textrm{non-SUSY)}/\sqrt\sigma=\sqrt{3/(2\pi)}\approx 0.7$. Correspondingly, in ${\cal N}=1$ SUSY YM, we conjecture that the Hagedorn temperature should be that of a non-critical (\textit{i.e.} well-defined in a 4-dimensional spacetime) closed superstring theory. Such a theory has been studied in particular in \cite{susystr}, where the usual Hagedorn temperature is recovered for the bosonic case and where the ratio 
\begin{equation}\label{TC}
\frac{T_c(\textrm{SUSY})}{T_c(\textrm{non-SUSY})}=\sqrt{\frac{2}{3}}\approx 0.8
\end{equation}
is found for the superstring. Interestingly the same value has been recently found in a SU(2) lattice simulation of ${\cal N}=1$ SUSY YM thermodynamics \cite{Bergner:2014saa}. Equation~(\ref{TC}) thus provides an explanation to this value, finally leading us to set $T_c(\textrm{SUSY})/\sqrt\sigma=1/\sqrt{\pi}\approx 0.6$. In \cite{LSCB,SUSYlong}, the ratio $T_c/\sqrt{\sigma} = 0.72$ has been chosen ($T_c=0.3$~GeV). In this paper, we take $T_c/\sqrt{\sigma} = 0.6$ ($T_c=0.25$~GeV). 

The gluon bare mass value $m_0/\sqrt \sigma=1.67$ were found by matching our $\cal T$-matrix results and the lattice ones in the bound state sector at $T=0$ of the non-SUSY YM case with gauge group SU(3) \cite{panero}. With this value, our model and the lattice data of \cite{panero} are in good agreement as shown in \cite{LSCB}. Because of supersymmetry, we further equal the gluino and gluon bare masses. This ratio $m_0/\sqrt \sigma$ is kept for any gauge group since all the dependence of the masses on the gauge group is assumed to come from the definition (\ref{m2}). This assumption is coherent with the lattice study \cite{Maasglu}, where the gluon propagator in Landau gauge has been shown to be nearly independent of the gauge group once normalized to the string tension. Let us now mention our results.

\section{Bound states}\label{secbound}
Bound states appear in our formalism as zeros of the Fredholm determinant with an energy below the threshold. The color singlet is the channel for which interactions are maximally attractive. Our computations show that color singlet two-body bound states can be formed above $T_c$; results are displayed in Table~\ref{tab1} for the most strongly bound states.

\begin{table}[h]
\caption{Masses (in units of $\sqrt\sigma$) of some bound states above $T_c$. A line mark the temperature at which a bound state is not detected anymore.}
\begin{tabular}{cccc}
\hline\hline
$T/T_c$ & $0^{-+}$ ($\tilde g\tilde g$) & $1/2$ ($\tilde g g$) & $0^{++}$ ($gg$) \\
\hline
1.05 & 3.10 & 4.53 & 4.53  \\ 
1.10 & 3.98 & 4.58 & 4.55 \\
1.15 & 4.10 & 4.46 & 4.43 \\ 
1.20 & 4.12 & 4.34 & 4.29 \\ 
1.25 & 4.07 & 4.24 & 4.22  \\
1.30 & 4.07 &  -   & - \\
1.35 & 4.05 &  \   & \  \\
1.40 & -    &  \   & \ \\
\hline\hline
\end{tabular}
\label{tab1} 
\end{table}

Two states are bound up to 1.30 $T_c$: $gg$ in $0^{++}$, which is the scalar glueball, and $g\tilde g$ in $J=1/2$. A $\tilde g \tilde g$ in $0^{-+}$, which is also called the adjoint $\eta'$ in the literature, can even be bound up to 1.40 $T_c$. Those results are valid for any gauge group in our formalism since $V(r,T)=- v(r,T)$ for two adjoint quasiparticles in the singlet representation. Although the existence of the adjoint $\eta'$ above $T_c$ is, to our knowledge, pointed out here for the first time, it is worth mentioning that the existence of the scalar glueball above $T_c$ and the decreasing trend of its mass we observe are compatible with previous lattice results \cite{Meng}.

We also find other bound states like tensor and pseudoscalar glueballs for example, as well as colored states in the adjoint representation; but they quickly dissolve and are not present anymore above typically 1.1 $T_c$. We note finally that the states appearing in Table~\ref{tab1} still appear as clear poles (resonances) of the $\cal T$-matrix  above the threshold up to around 1.5 $T_c$.

\section{Equation of state}\label{seceos}
\begin{figure}[h!]
\begin{center}
\includegraphics*[width=0.7\textwidth]{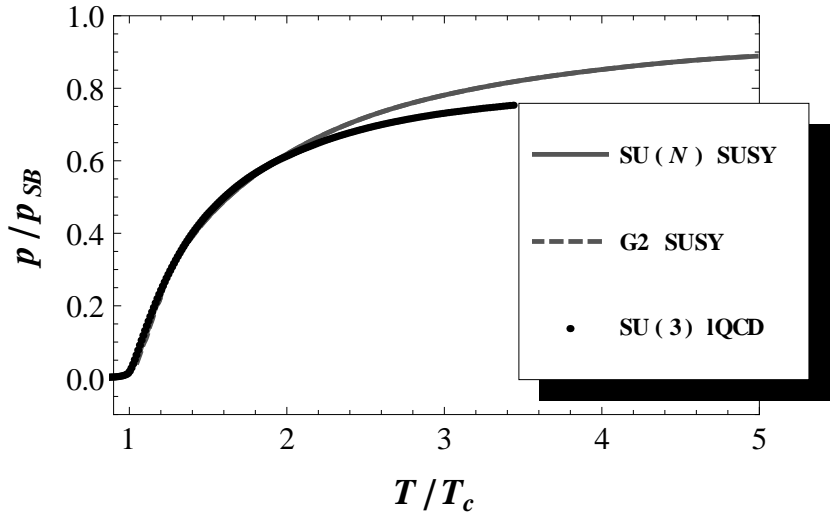}
\caption{Normalized pressures $p/p_{SB}$ versus $T/T_c$, computed for ${\cal N}=1$ SUSY YM, with gauge groups SU(2), SU(3) and SU($\infty$) commonly denoted as SU($N$) (solid gray line) and G$_2$ (dashed gray line). The pressure for non-SUSY YM with gauge group SU(3) is shown for comparison (black dots); data are taken from the lattice study \cite{panero}.}
\label{pressure}
\end{center}
\end{figure}

Thanks to Eqs.~(\ref{pot0}) and (\ref{pta}), the pressure and trace anomaly of the ${\cal N}=1$ SUSY YM theory can be computed. They are normalized to the Stefan-Boltzman pressure $p_{SB}=\pi^2 T^4 {\dim}\ adj\, /24$. We focus mainly on the temperature range (1-5)~$T_c$ in which the medium is presumably in a strongly coupled phase. The pressure is plotted in Fig.~\ref{pressure} for all the gauge groups investigated: SU(2), SU(3), SU($\infty$) and G$_2$. All those curves are indistinguishable, showing a very weak dependence of the pressure on the gauge group in our model. The SU(3) pressure computed in the non-SUSY case on the lattice is shown. This curve is surprisingly close to the newly computed EoS.

\begin{figure}[h!]
\begin{center}
\includegraphics*[width=0.7\textwidth]{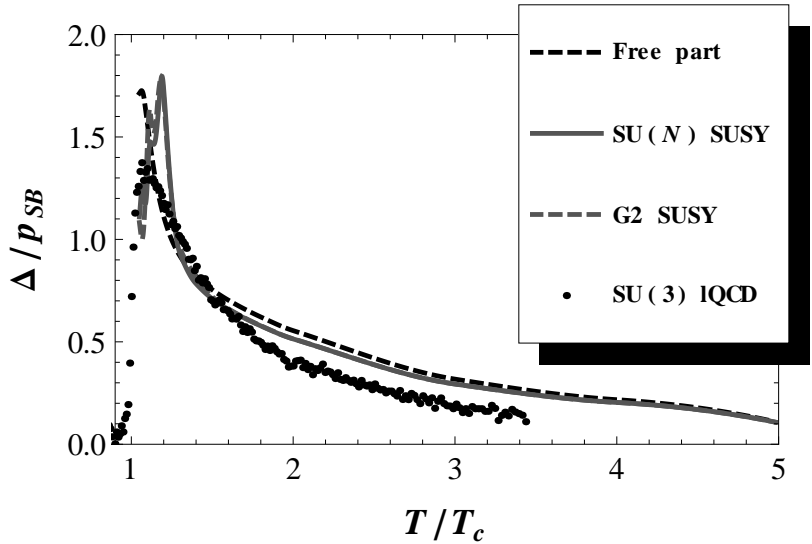}
\end{center}
\caption{Same as Fig.~\ref{pressure} for the normalized trace anomaly $\Delta/p_{SB}$. The free part of the trace anomaly is shown for comparison (dashed black line) and is gauge-group independent.}
\label{traceanom}
\end{figure}

The trace anomaly is then displayed in Fig.~\ref{traceanom} for the same gauge groups, and the same conclusions as for the pressure apply. In all cases the trace anomaly has a peak around 1.2 $T_c$, which is actually the temperature at which the bound-state and scattering contributions to the grand potential are maximal \cite{SUSYlong}. It is worth stressing that this peak structure is really due to two-body interactions because it is absent in the free gas contribution, also displayed in Fig.~\ref{traceanom}. As for the pressure, the curves obtained for SU($N$) are rather close to the previously known non-SUSY ones.

At higher temperatures, the interaction potential progressively vanishes and the pressure tends toward its Stefan-Boltzman value while the trace anomaly correspondingly tends toward zero. As the potential becomes weak enough, our model is accurately described within the Born approximation ${\cal T}=V+\,{\rm O}(V^2)$. In this case, the color structure of the two-body interactions leads to the vanishing of gluino-gluon interactions in the medium: Gluons and gluinos do not interact with each other at high temperature in average. However, interactions between gluons only and gluinos only are still present \cite{SUSYlong}.

\begin{figure}[h!]
\begin{center}
\includegraphics*[width=0.7\textwidth]{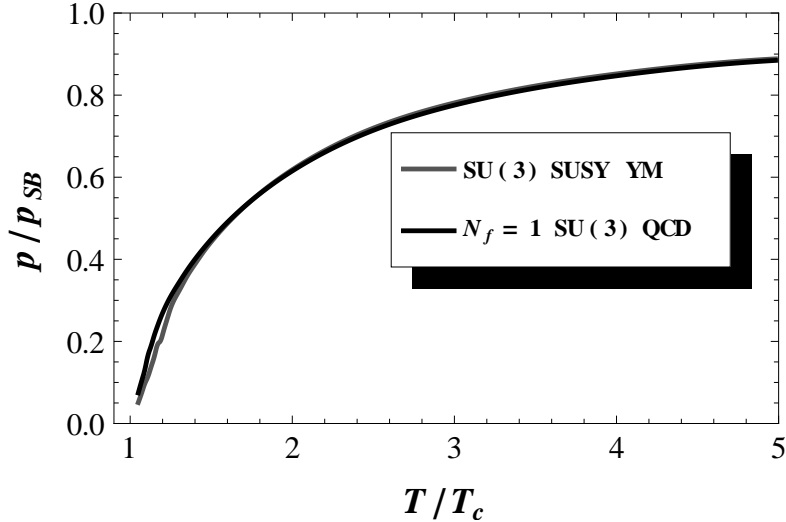}
\caption{Normalized pressure $p/p_{SB}$ versus $T/T_c$, computed for SU(3) one-flavor QCD (black line) and ${\cal N}=1$ SU(3) SUSY YM (gray line). Each case is normalized to its own Stefan-Boltzmann pressure. 
}
\label{Orient1}
\end{center}
\end{figure}

The study of the EoS can shed some light on the orientifold equivalence too. It states that a SU($N$) YM theory with $N_f$  Dirac fermions in the two-index antisymmetric color representation and a SU($N$) YM theory with $N_f$ Majorana flavors in the adjoint representation are equivalent at large $N$ in the bosonic sector \cite{Armo03}. When $N_f = 1$, this equivalence relates $\mathcal{N} = 1$ SUSY YM and the so-called $N_f=1$ QCD$_{\textrm{AS}}$, reducing to standard $N_f=1$ QCD for $N=3$. As shown in detail in \cite{SUSYlong}, the orientifold equivalence holds at the level of the EoS within our formalism. Moreover, we are able to study how different both two theories are at $N=3$. The normalized pressures, plotted in Fig.~\ref{Orient1}, are almost identical although some differences can be found very close to $T_c$. It appears that SU(3) $\mathcal{N} = 1$ SUSY YM provides a good approximation of SU(3) $N_f=1$ QCD at the level of the EoS.

\section{Concluding comments}\label{secconclu}
We have studied for the first time the properties of the deconfined phase of the ${\cal N}=1$ SUSY YM by resorting to a ${\cal T}$-matrix formulation. We have shown that the 20\% decrease of $T_c$, observed on the lattice when adding SUSY \cite{Bergner:2014saa}, is compatible with the identification of $T_c$ to the Hagedorn temperature of a non-critical closed superstring. This leads us to the estimate $T_c=250$~MeV, which should be close to the deconfinement temperature of one-flavor QCD in virtue of the orientifold equivalence. The normalized EoS shows almost no dependence on the gauge group. Although we focused on SU($N$) and G$_2$ in our numerical calculations, the EoS is not expected to be strongly different for the other cases because one always has $adj\otimes adj=\bullet\oplus adj^A\oplus \textrm{higher\ dim}$. The singlet and antisymmetric adjoint representations generate bound states and attractive interactions, while the higher dimensional representations generally generate repulsive interactions. Their number and associated color factors differ from one gauge group to another, but qualitatively influence the EoS in the same way. Moreover, the EoS is found to be very close to the non-SUSY EoS as computed in lattice computation. Finally, two-body bound states can exist up to 1.4 $T_c$, around the peak of the trace anomaly. The recent progress made in handling supersymmetry on a lattice let us hope that these results may be compared to lattice calculations in a near future.

\textit{Acknowledgments -- } G.L. thanks the F.R.S-FNRS for financial support.


\begin{thebibliography}{99}
 \bibitem{yagi} K. Yagi, T. Hatsuda, and Y. Miake, \textit{Quark-Gluon Plasma: From Big Bang to Little Bang} (Cambridge Monographs on Particle Physics, Nuclear Physics and Cosmology, Cambridge University Press, 2008).
\bibitem{G2} M.~Pepe and U.-J.~Wiese, Nucl.\ Phys.\ B {\bf 768}, 21 (2007).
\bibitem{braun} J.~Braun, A.~Eichhorn, H.~Gies, and J.~M.~Pawlowski, Eur.\ Phys.\ J.\ C {\bf 70}, 689 (2010).
\bibitem{Dumitru:2012fw} 
  A.~Dumitru, Y.~Guo, Y.~Hidaka, C.~P.~Korthals~Altes, and R.~D.~Pisarski,
  Phys.\ Rev.\ D {\bf 86}, 105017 (2012).
  \bibitem{LSCB} G. Lacroix, C. Semay, D. Cabrera, and F. Buisseret, Phys.\ Rev.\  D {\bf 87}, 054025 (2013).
    \bibitem{panero} 
      M.~Panero,
      Phys.\ Rev.\ Lett.\  {\bf 103}, 232001 (2009).
\bibitem{salam} 
  A.~Salam and J.~A.~Strathdee,
  Phys.\ Lett.\ B {\bf 51}, 353 (1974).
\bibitem{Novi83} 
  V.~A.~Novikov, M.~A.~Shifman, A.~I.~Vainshtein, and V.~I.~Zakharov,
  Nucl.\ Phys.\ B {\bf 229}, 381 (1983).
\bibitem{Farrar:1997fn}
  G.~Veneziano and S.~Yankielowicz,
  Phys.\ Lett.\ B {\bf 113}, 231 (1982); 
   A.~Feo, P.~Merlatti, and F.~Sannino,
    Phys.\ Rev.\ D {\bf 70}, 096004 (2004).
\bibitem{Bergner:2013nwa}
  G.~Bergner, I.~Montvay, G.~M\"unster, U.~D.~\"Ozugurel, and D.~Sandbrink,
  JHEP {\bf 1311}, 061 (2013).
\bibitem{Bergner:2014saa} 
  G.~Bergner, P.~Giudice, G.~M\"unster, S.~Piemonte, and D.~Sandbrink,
  arXiv:1405.3180.
\bibitem{Anber:2014lba} 
  M.~M.~Anber, E.~Poppitz, and B.~Teeple,
  arXiv:1406.1199.
\bibitem{Amat88} 
  D.~Amati, K.~Konishi, Y.~Meurice, G.~C.~Rossi, and G.~Veneziano,
  Phys.\ Rep.\  {\bf 162}, 169 (1988).
\bibitem{Armo03} 
  A.~Armoni, M.~Shifman, and G.~Veneziano,
  Phys.\ Rev.\ Lett.\  {\bf 91}, 191601 (2003).
\bibitem{SUSYlong} G. Lacroix, C. Semay, and F. Buisseret, arXiv:1408.0958. 
\bibitem{Shuryak:2004tx} 
  E.~V.~Shuryak and I.~Zahed,
  Phys.\ Rev.\ D {\bf 70}, 054507 (2004).
  \bibitem{tmat} D.~Cabrera and R.~Rapp,
    Eur.\ Phys.\ J.\ A {\bf 31}, 858 (2007); Phys.\ Rev.\  D {\bf 76}, 114506 (2007).
\bibitem{HTL} J.-P. Blaizot, E. Iancu, and A. Rebhan, Phys. Lett. B \textbf{470},
181 (1999); Phys. Rev. D \textbf{63}, 065003 (2001).
\bibitem{dashen} R. Dashen, S.-K. Ma, and H. J. Bernstein, Phys. Rev. \textbf{187}, 345 (1969).
\bibitem{haftel} M.~I. Haftel and F. Tabakin, Nucl. Phys. A \textbf{158}, 1 (1970).
\bibitem{Prat94} 
  S.~Pratt and W.~Bauer,
  Phys.\ Lett.\ B {\bf 329}, 413 (1994).
\bibitem{jaco} M.~Jacob and G.~C.~Wick, Annals Phys.\  {\bf 7}, 404 (1959).
\bibitem{latpot} O. Kaczmarek, F. Karsch, P. Petreczky, and F. Zantow,
Phys. Lett. B \textbf{543}, 41 (2002).
\bibitem{hage} H. B. Meyer, Phys. Rev. D 80, 051502(R) (2009); 
      M. Caselle, L. Castagnini, A. Feo, F. Gliozzi, and M. Panero,
      JHEP {\bf 1106}, 142 (2011).
   \bibitem{hage2} 
  F.~Buisseret and G.~Lacroix,
  Phys.\ Lett.\ B {\bf 705}, 405 (2011).
  \bibitem{susystr} A. H. Chamseddine, Phys. Lett. B \textbf{258}, 97 (1991);
    S.~D.~Odintsov,
    Phys.\ Lett.\ B {\bf 274}, 338 (1992).
      \bibitem{Maasglu} 
        A.~Maas,
        JHEP {\bf 1102}, 076 (2011).
\bibitem{Meng} 
  X.~-F.~Meng \textit{et al.} [CLQCD Collaboration],
  Phys.\ Rev.\ D {\bf 80}, 114502 (2009).

\end{thebibliography}
\end{document}